\documentclass[%
 aip,
 amsmath,amssymb,
 reprint,%
]{revtex4-1}
\usepackage{physics}
\usepackage{graphicx}
\usepackage[usenames,dvipsnames,svgnames,table]{xcolor}
\usepackage{tcolorbox}
\usepackage{amsmath} 
\usepackage{amsfonts} 
\usepackage{amssymb} 
\usepackage{subfigure}
\usepackage{bm}
\usepackage{booktabs}   
\usepackage{siunitx}    
\usepackage{upgreek}
\usepackage[detect-all]{siunitx}
\usepackage[unicode=true,
pdfusetitle,
bookmarks=true,
bookmarksnumbered=false,
bookmarksopen=false,
breaklinks=true,
pdfborder={0 0 0},
backref=false,
colorlinks=true,
hypertexnames=false]{hyperref}
\hypersetup{linkcolor=NavyBlue,urlcolor=NavyBlue,citecolor=NavyBlue}
\usepackage[english]{babel} 
\usepackage{bbm}
\usepackage{graphicx,color}

\definecolor{mygreen}{rgb}{0,0.5,0}
\definecolor{myred}{rgb}{0.75,0,0}
\definecolor{myblue}{rgb}{0,0,0.75}

\newcommand{\myaffiliation}{\affiliation}
\newcommand{\HDU}
{\myaffiliation{Department of Physics, Hangzhou Dianzi University, Hangzhou 310018, China}}
\newcommand{\ZKL}
{\myaffiliation{Zhejiang Key Laboratory of Quantum State Control and Optical Field Manipulation, Hangzhou Dianzi University, Hangzhou 310018, China}}

\begin{document}
\title{All-Optical High-Resolution Real-Time Temperature Estimation Method Based on Fiber-Optic Interferometry\\
}
\author{Jingwen Yang}
\thanks{These authors contributed equally to this work.}
\HDU

\author{Long Chen}
\thanks{These authors contributed equally to this work.}
\HDU

\author{Haoliang Yu}
\HDU

\author{Xiaofeng Jin}
\HDU

\author{Jianxiang Miao}
\HDU

\author{Jia Kong}
\email{jia.kong@hdu.edu.cn}
\HDU
\ZKL






\date{\today}

\begin{abstract}

High-resolution temperature monitoring is essential for many engineering and scientific applications, but conventional sensors are limited by insufficient resolution and susceptibility to electromagnetic interference. Fiber-optic interferometers provide high sensitivity and intrinsic electromagnetic immunity; however, their practical performance is hindered by nonlinear temperature–intensity responses, phase ambiguity, and environmental disturbances. Here, we develop an extended Kalman filter (EKF)-based approach that incorporates system nonlinearity and noise statistics to enable robust real-time temperature estimation from interferometric signals. In numerical simulations, our EKF-based method reduces the estimation error to $\SI{2.21e-5}{\kelvin}$, while experiments achieve a resolution of $\SI{8.34e-5}{\kelvin}$ under strong disturbances, corresponding to a threefold improvement over conventional intensity-based inversion method and an order-of-magnitude enhancement compared with thermistor-based measurement.
These results demonstrate a compact and robust strategy for high-resolution, real-time, all-optical temperature sensing with strong immunity to electromagnetic interference. 

\end{abstract}

\maketitle

\section{Introduction}
	
Temperature sensing is essential in a wide range of applications, including industrial production\cite{Nedelcu2024, Xu2025}, environmental monitoring\cite{Jin2025, Short2003}, medical diagnostics\cite{Cui2005, Schena2016, Ta2023}, energy systems\cite{Chen2025, Wang2021}, food processing\cite{Cao2025}, and aerospace engineering\cite{Tonks2006}. High-precision applications require sensors with high temperature resolution to ensure measurement accuracy\cite{Li2022, Sheikh2009}. Common temperature sensors include NTC thermistors, platinum resistance thermometers (PT100/PT1000), thermocouples, digital sensors (e.g., DS18B20), and infrared sensors. Although each has its own advantages, their temperature resolution is generally limited: NTC thermistors achieve ~$\SI{e-3}{\kelvin}$ over a narrow range ($\SI{-50}{\degreeCelsius}$–$\SI{150}{\degreeCelsius}$), platinum resistance thermometers offer $\SIrange{e-3}{e-2}{\kelvin}$ over $\SI{-200}{\degreeCelsius}$–$\SI{600}{\degreeCelsius}$\cite{Svelto2001}, thermocouples provide a wider temperature range ($\SI{-270}{\degreeCelsius}$–$\SI{1800}{\degreeCelsius}$) but lower resolution ($\SI{e-2}{\kelvin}$)\cite{article}, and digital sensors are typically limited to ~$\SI{0.1}{\kelvin}$\cite{Runjing2011}. Infrared sensors enable non-contact measurement but are susceptibile to emissivity and environmental conditions\cite{Yoo2011}.

In contrast, optical fiber interferometers\cite{Pang2011, Abbas2021} detect temperature through phase-induced intensity variations, offering high susceptible and intrinsic immunity to electromagnetic interference\cite{Liu2025, Xiong2025}. However, conventional intensity-based inversion methods, which infer temperature directly from interference intensity, rely on a nonlinear temperature–intensity relationship, and their resolution is significantly degraded by laser fluctuations, airflow, mechanical vibrations, and detector noise, which limits their performance in high-resolution applications.
These disturbances become particularly critical in high-resolution temperature measurements, where even minor noise can significantly degrade estimation accuracy. The Kalman filter (KF)\cite{Kalman1960} is a well-established approach for dynamic signal tracking and noise suppression. By exploiting the statistical properties of process and measurement noise, it effectively reconstructs the underlying signal from noisy observations. For linear Gaussian systems, the KF provides optimal real-time estimation with recursive updating capability\cite{book, Kalman1960}. 
However, temperature sensing based on interferometric intensity signals typically involves inherently nonlinear observation models. Therefore, the extended Kalman filter (EKF) \cite{welch2001introduction, SUNAHARA1970}, a nonlinear variant of the standard KF, is required to linearize the nonlinear measurement function around the current estimate.
In this work, we propose an all-optical temperature sensing method based on a Mach–Zehnder (MZ) fiber interferometer\cite{Chen1989, Lee2011, Kong2015, Zhu2024}, in which the EKF is applied to enable high-resolution, real-time temperature estimation. The proposed method is benchmarked against conventional intensity-based inversion and thermistor-based measurements. Both simulation and experimental results demonstrate that the EKF approach provides stable real-time tracking, and significantly enhances temperature resolution while effectively suppressing measurement noise. As a result, this study offers a robust framework for improving the performance and reliability of high-resolution optical temperature sensing systems.
	
\section{Principles}

\begin{figure} [b]
	\centering
	\includegraphics[width=0.85\linewidth]{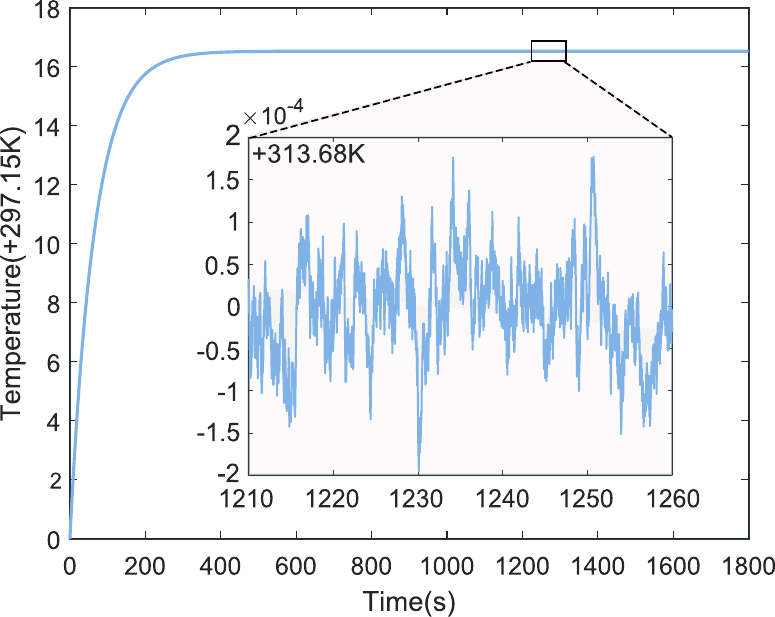}
	\caption{
    Simulation of a dynamic heating process. The initial temperature is set to $\SI{297.15}{\kelvin}$, the steady-state temperature to $\SI{313.68}{\kelvin}$, the thermal response coefficient to $\SI{0.01531}{\per\second}$, and the heating duration is 1800 s.}
	\label{fig:Temperature}
\end{figure}

To evaluate the feasibility and performance of the proposed estimation framework, we consider a representative dynamic temperature sensing scenario based on a heating process, as commonly adopted in related studies \cite{Vollmer2009, Faraoni2020}. The temperature evolution during heating is modeled using an exponential function:
\begin{equation}
	T(t) = T_\infty - \left(T_\infty - T_0\right) \cdot e^{-kt}+\tau (t),
	\label{heating}
\end{equation}
where $T_0$ and $T_\infty$ denote the initial and steady-state temperatures, respectively, $k$ is the thermal response coefficient determined by sensor and environmental properties, and $\tau(t)$ represents stochastic temperature fluctuations. To more realistically capture environmental fluctuations, the fluctuation term $\tau(t)$ is modeled as an Ornstein–Uhlenbeck (OU) process \cite{CheTaib2025}:
\begin{equation}
	d\tau(t) = \theta (\mu - \tau(t)) dt + \sigma dW(t),
	\label{ou}
\end{equation}
where $\theta>0$ characterizes the reversion rate toward the equilibrium value $\mu$, and $\sigma>0$ defines the noise intensity driven by the Wiener process $dW(t)$ with $E[dW(t)^2]=dt$. In numerical implementation, the OU process is discretized with time step $dt$ and parameterized by $\theta=2~\mathrm{s}^{-1}$ and $\sigma=1\times10^{-4}~\mathrm{K\cdot s^{-1/2}}$. The resulting stochastic fluctuations are superimposed onto the deterministic thermal model in Eq.~\eqref{heating}, yielding temperature variations with an root-mean-square error (RMSE) of $5.07\times 10^{-5}\text{K}$, as illustrated in Fig.~\ref{fig:Temperature}.
In practical sensing systems, such subtle fluctuations are difficult to resolve using conventional approaches. Due to circuit limitations, environmental disturbances, and self-heating effects \cite{Liu2024}, even precision devices such as 10 k$\Omega$ thermistors often fail to achieve the required resolution, motivating the need for more robust dynamic estimation strategies. 

\begin{figure} [t]
	\centering
	\includegraphics[width=1\linewidth]{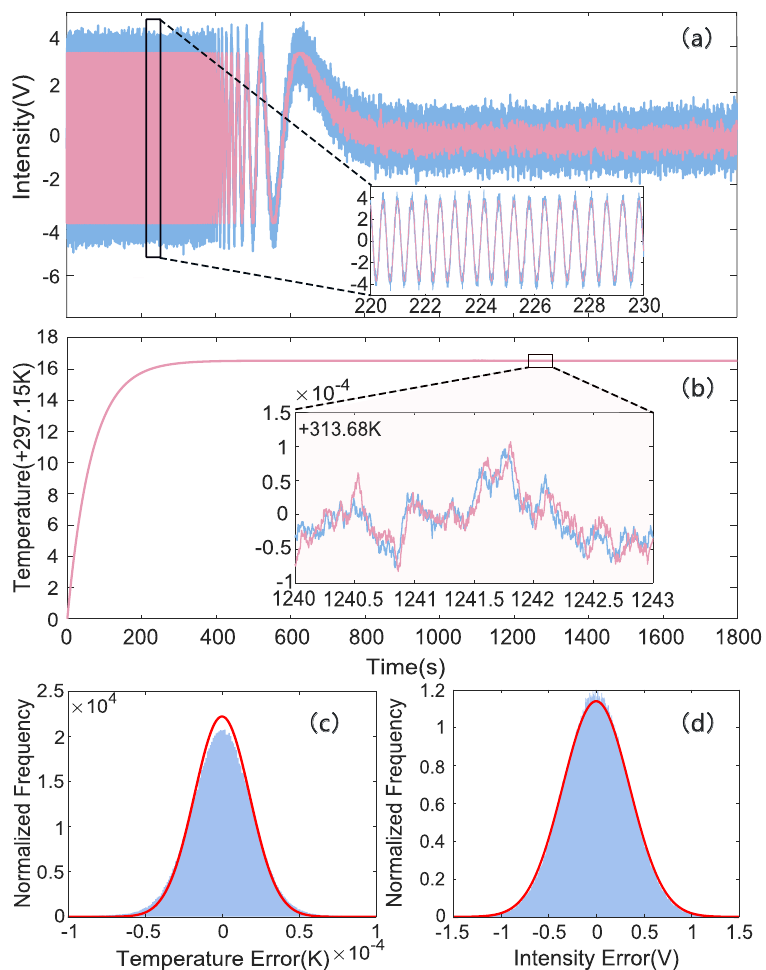}
	\caption{From non-steady to steady-state  simulation and estimation results. (a) Simulated (blue curve) and EKF-estimated light intensity (pink curve). (b) Simulated (blue curve) and EKF-estimated temperature (pink curve). (c) Histogram of the normalized light intensity error distribution with the corresponding Gaussian curve defined by the innovation covariance.(d) Histogram of the normalized temperature error distribution with the corresponding Gaussian curve defined by the posterior covariance matrix.}
	\label{fig:simulation}
\end{figure}

In a MZ fiber interferometer, these temperature variations induce phase shifts that are ultimately converted into measurable optical intensity fluctuations. The relationship between light intensity and temperature can be expressed as:
\begin{equation}
	I(t) = 2I_0 \cos\left( \frac{\mathrm{d}\varphi}{\mathrm{d}T} (T(t) - T_0) + \varphi_0 \right),
\label{intensity_eq}
\end{equation}
where $I_0$ is the amplitude of light intensity, $\dfrac{\mathrm{d}\varphi}{\mathrm{d}T}$ is the temperature-induced phase response and $\varphi_0$ is initial phase offset. By detecting the light intensity signal $I(t)$, the corresponding temperature change $T(t) - T_0$ can be inferred through Eq.~\eqref{intensity_eq}. However, in practice, the signal is often corrupted by laser intensity fluctuations, environmental airflow, mechanical vibrations, and optical shot noise, which severely degrade the accuracy of conventional intensity-based inversion methods. 

From a signal processing perspective, this problem can be naturally formulated as a nonlinear state estimation problem, where temperature evolves as the hidden state and optical intensity serves as the noisy nonlinear observation. It is therefore necessary to construct a state–space representation and apply recursive estimation techniques capable of handling both noise and nonlinearity.
It is important to note that the temperature–intensity mapping is inherently nonlinear, making conventional linear KF unsuitable. To address this issue, the EKF is employed as a nonlinear extension of the standard KF. By locally linearizing the nonlinear observation model, the EKF enables recursive estimation of the system state under known noise statistics, effectively suppressing measurement disturbances and improving the tracking of small, rapid temperature variations. Accordingly, for this nonlinear system with state $\mathbf{x}_n$ (temperatures) and observation ${z}_n$ (light intensity), the EKF is adopted for state estimation, and the system is described by the following state-space model:
\begin{equation}
	\mathbf{x}_n =\mathbf{\Phi}  \mathbf{x}_{n-1}+{\mathbf{\omega}}_{n},
	\label{state}
\end{equation}
\begin{equation}
	 z_{n} =  2I_0 \cos\left(\begin{bmatrix} \frac{\mathrm{d}\varphi}{\mathrm{d}T}
		 &0 \end{bmatrix} (\mathbf{x}_n-T_{0}) + \varphi_0 \right)
          +\nu  _{n} .
\label{observation}
\end{equation}
Here $\mathbf{\Phi}=\begin{bmatrix}
		\frac{1}{1+kdt }& \frac{kdt}{1+kdt } \\
			0 &1
		\end{bmatrix}$, $\mathbf{x}_n=\mathbf{\begin{bmatrix}
	   T_{n}&T_{\infty} 
	\end{bmatrix}}^T$, $T_{n}$ is real time temperature, $T_{\infty}$ is steady-state temperature, $dt = t_n - t_{n-1}$ is the discrete-time step and $n$ is an integer. The process noise $\mathbf{\omega}_{n} \sim \mathcal{N}(0, \mathbf{Q})$ and the observation noise $\nu_{n} \sim \mathcal{N}(0, R)$ are both assumed to be Gaussian distributions, with $\mathbf{Q}$ and $R$ being their respective covariance matrices.

In this system, the initial temperature $T_0$ and the thermal response coefficient $k$ are assumed to be known, while the steady-state temperature $T_\infty$ is treated as an unknown parameter to be estimated. The temperature at each time step is regarded as the system state to be reconstructed.
  At time $t_{n}$, the prior state estimate $\hat {\mathbf{x}}_{n|n-1}$ and the prediction error covariance $\mathbf{P}_{n|n-1} $ are used to make predictions based on the model and the previous posterior estimate $\hat{\mathbf{x}}_{n-1|n-1}$. They are given by
\begin{equation}
	\hat {\mathbf{x}}_{n|n-1} = \mathbf{\Phi}  {\hat{\mathbf {x}}}_{n-1|n-1},
\end{equation}
\begin{equation}
		\mathbf{P}_{n|n-1} = \mathbf F \mathbf{P}_{n-1|n-1} \mathbf {F }^T+ \mathbf{Q}.
\end{equation}
Here $\mathbf{F}$ denotes the Jacobian matrix of the state transition function. In this case, it reduces to $\mathbf{F}=\mathbf{\Phi}$.

During the update step, the state estimate is refined by incorporating the actual measurement $z_{n}$ together with the predicted observation $\hat{ z}{_{n} }$. The posterior state estimate $\hat {{T}}_{n|n}$ and its covariance matrix $\mathbf {P}_{n|n}$ are updated as:
\begin{equation}
	\hat {\mathbf{x}}_{n|n} = \hat {\mathbf{x}}_{n|n-1}+ \mathbf{K}_n ( z_{n}-\hat{ z}{_{n} }),
\end{equation}
\begin{equation}
	\mathbf{P}_{n|n} = \left( \mathbf{I} - \mathbf{K}_n \mathbf{H}_n \right) \mathbf{P}_{n|n-1},
	\label{P}
\end{equation}
where $\hat{\mathbf{x}}_{n|n}=\mathbf{\begin{bmatrix}
	   \hat{T}_{n|n}&\hat{T}_{\infty n|n} 
	\end{bmatrix}}^T$, $\hat{T}_{n|n}$ is posterior real time temperature estimate, $\hat{T}_{\infty n|n}$ is posterior steady-state temperature, while $\hat{z}{_{n} }=2I_0 \cos\left(\frac{\mathrm{d}\varphi}{\mathrm{d}T}
		  (\hat {{T}}_{n|n-1}-T_{0}) + \varphi_0 \right)$ denotes the predicted observation. The observation matrix $\mathbf H_{n}$ is obtained from the first-order Taylor expansion of Eq.~\eqref{observation}, given by
$\mathbf H_{n} = \begin{bmatrix}-2I_0 \frac{\mathrm{d}\varphi}{{d}T}
\sin\left( \frac{\mathrm{d}\varphi}{\mathrm{d}T} (\hat {{T}}_{n|n-1} - T_0) + \varphi_0 \right) & 0 \end{bmatrix}$.
Here $\mathbf I$ is the identity matrix, and $\mathbf {K}_{n}$ is the Kalman gain defined as:
\begin{equation}
	  \mathbf{{K}_{n}}=\mathbf{P}_{n|n-1}\mathbf{H}_n^T {S}_n^{-1},
\end{equation}
which determines the weighting between prediction and measurement in the update process. The innovation covariance matrix ${S}_n$ is given by
\begin{equation}
	{S}_n = E[(z_{n}-\hat{z}{_{n} } )({ z_{n}-\hat{z}{_{n} }})^T]=\mathbf{H}_n \mathbf{P}_{n|n-1} \mathbf{H}_n^T + {R}.
	\label{S}
\end{equation}
where ${R}$ determined by the power spectral density of the observation ${z_{n}}$. 

The simulation data are generated using parameters selected to match the experimental conditions, with $T_0=297.15\text{K}$, $T_\infty=313.68\text{K}$, and $k={0.01531}\text{K}\cdot s^{-1}$. The EKF is then applied to the simulated data, where $T_n$ and $T_\infty$ are treated as quantities to be estimated, while only $T_0$ and thermal response coefficient are assumed to be known. The simulated and estimated intensity signals are shown in Fig.~\ref{fig:simulation}(a), represented by the blue and pink curves, respectively. To evaluate the estimation performance, the true temperature evolution and the EKF estimation are compared in Fig.~\ref{fig:simulation}(b). The enlarged view shows that the EKF estimation closely follows the temperature dynamics.
 These results indicate that the EKF can accurately reconstruct the underlying temperature profile.

To further validate the model consistency and EKF implementation, the estimation errors ${{T}}_n-\hat{{T}}_{n|n}$ and ${z_{n}}-\hat{{z}}{_{n} }$ are examined and shown to follow zero-mean Gaussian distributions with steady state variances given in Eqs.~\eqref{P} and ~\eqref{S}, respectively. 
The corresponding histograms in Figs.~\ref{fig:simulation}(c) and (d) are in good agreement with the theoretical Gaussian distributions.  Specifically, $\SI{92.4}{\percent}$ and  $\SI{95.7}{\percent}$ of the data fall within the $\SI{95}{\percent}$ confidence intervals for ${{T}}_n-\hat{{T}}_{n|n}$ and ${z_{n}}-\hat{z}_{n}$, respectively, confirming strong consistency between the model and simulation results.
Additionally, the RMSE between the estimated and true temperature is calculated to be $\SI{2.21e-5}{\kelvin}$, demonstrating the capability of the proposed EKF method for high-resolution temperature sensing in practical applications.

\section{Experimental results}

\begin{figure*} [t]
    \centering
    \includegraphics[width=0.8\textwidth]{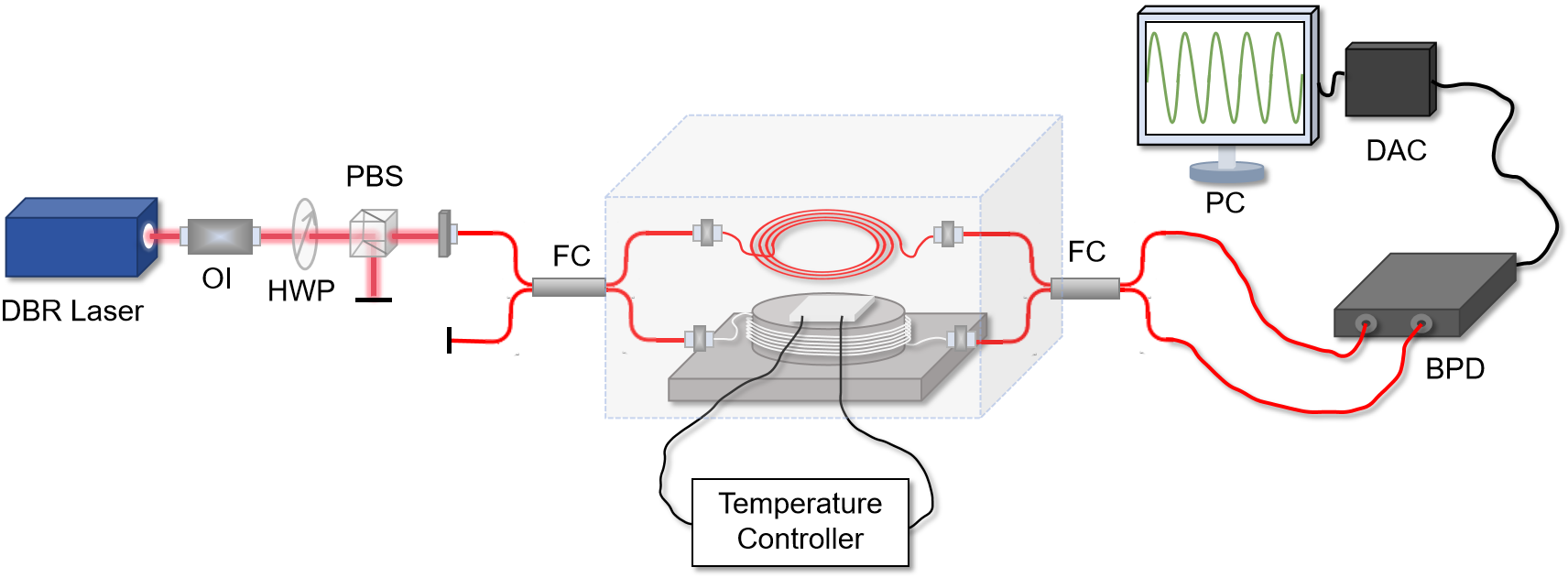}
    \caption{Schematic of the experimental setup. A distributed Bragg reflector (DBR) laser operating at ~795 nm is used as the light source. The beam passes through an optical isolator (OI), a half-wave plate (HWP), and a polarizing beam splitter (PBS), and is then divided into the sensing and reference arms by a 50:50 fiber coupler (FC). After propagating through both arms, the beams are recombined by a second FC, and the resulting interference signal is detected by a balanced photo-detector (BPD) and recorded using a data acquisition card (DAC) with a bandwidth of 100 kHz.}
    \label{fig:setup}
\end{figure*}

\begin{figure}[b]
	\centering
	\includegraphics[width=0.95\linewidth]{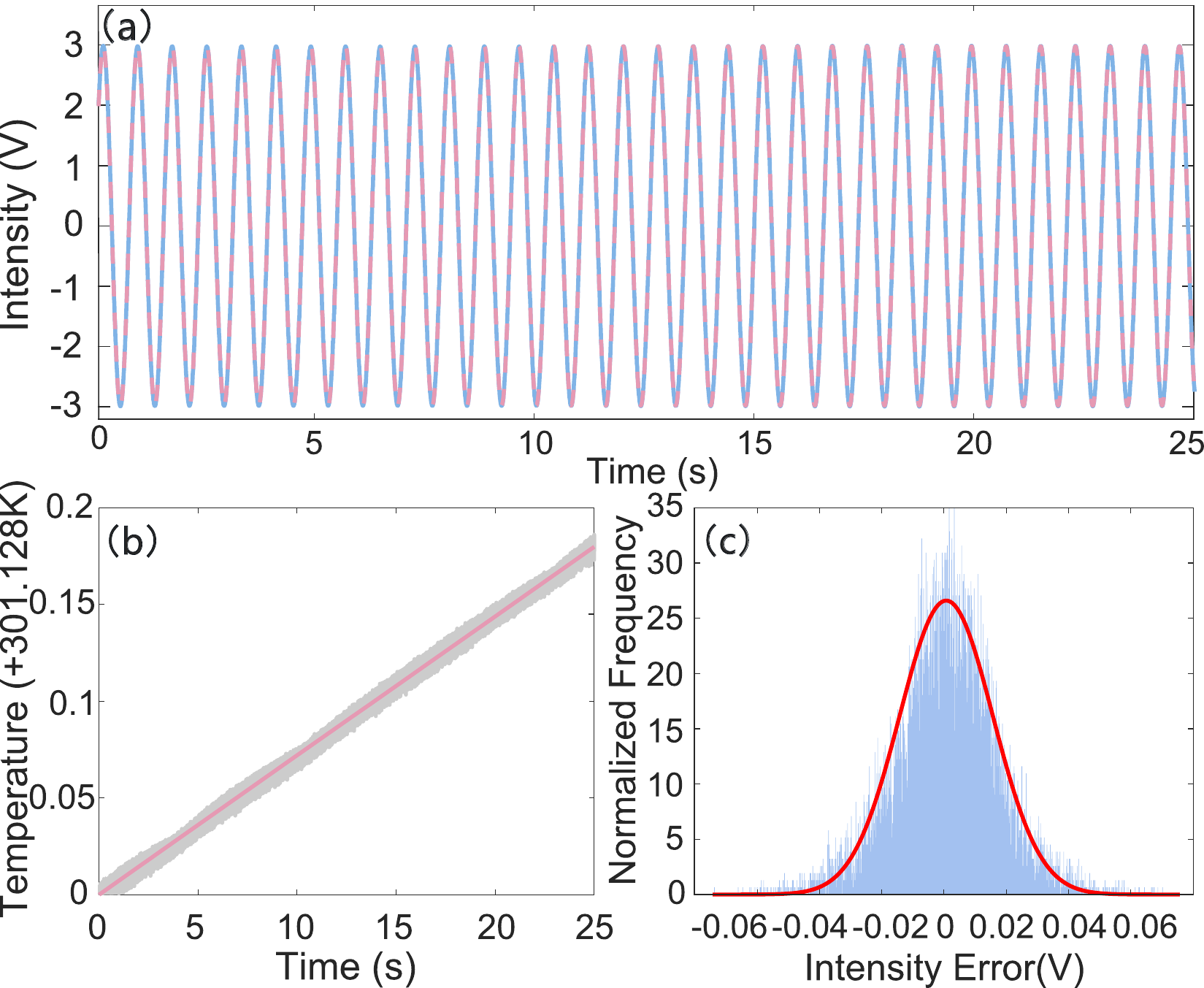}
	\caption{Heating process experimental results. (a) Measured light intensity (blue solid curve) and EKF estimation (pink dashed curve).	(b) Temperature monitoring using a 10 $\text{k}\Omega$ thermistor (gray curve) and EKF estimation (pink curve).
	(c)Histogram of the normalized light intensity error distribution with the corresponding Gaussian curve defined by the innovation covariance.}
	\label{fig:nonsteady}
\end{figure}

\begin{figure}[b]
	\centering
	\includegraphics[width=0.95\linewidth]{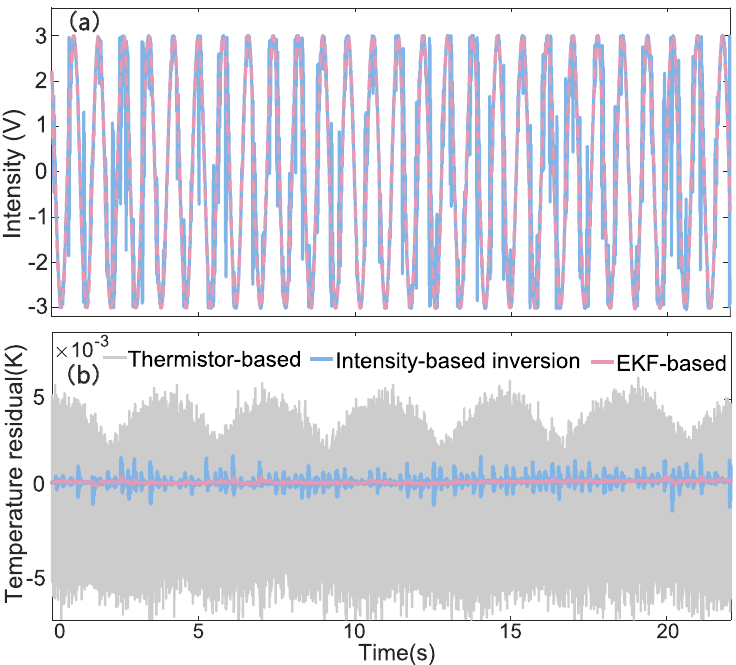}
	\caption{Experimental results under disturbance conditions. (a) Measured light intensity with noise (blue solid curve) and EKF estimation (pink dashed curve). (b) Comparison of temperature residuals obtained from the thermistor-based method, the conventional intensity-based inversion method, and the EKF-based method.}
	\label{fig:nonsteady_noise}
\end{figure}

The experimental configuration of the MZ fiber interferometer is illustrated in Fig.~\ref{fig:setup}. The sensing arm consists of 5 m of bare fiber with a 245 $\upmu$m coating, tightly hand-wound around an aluminum-alloy cylinder with a diameter of 10 cm at room temperature. In contrast, the reference arm comprises 5 m of jacketed fiber with a 900 $\upmu$m coating. A 10 k$\Omega$ thermistor is attached to the aluminum cylinder to monitor the system temperature. 
To prevent stress-induced polarization fluctuations, all fibers used in the interferometer are polarization-maintaining fibers. The entire interferometer assembly is enclosed within an insulating foam box (dimensions: 410×270×210 mm$^{3}$; wall thickness: 40 mm), to reduce external thermal perturbations. To suppress common-mode technical noise and enhance the signal, the differential signal between the two interference outputs is used, enabling precise measurement of temperature-induced phase shifts in the sensing fiber.

Based on this setup, under a nonstationary heating process, the EKF is applied to estimate real-time temperature dynamics from the measured light intensity, given a known initial temperature and thermal response coefficient. As shown in Fig.~\ref{fig:nonsteady}(a), the measured intensity (blue solid curve) and EKF estimate (pink dashed curve) agree closely. Fig.~\ref{fig:nonsteady}(b) compares the temperature measured by a 10 k$\Omega$ thermistor (gray) with the EKF estimate (pink). The EKF estimate closely tracks the thermistor-measured trend, exhibiting substantially smaller fluctuations, which supports its effectiveness for dynamic estimation. Owing to the limited accuracy of the thermistor, which cannot fully capture the actual temperature variations, direct comparison with the true temperature is not feasible. To further assess the reliability of the EKF estimates, a statistical analysis is performed by comparing the histogram of intensity residuals with the theoretical Gaussian distribution defined by Eq.~\eqref{S}. As shown in Fig.~\ref{fig:nonsteady}(c), $\SI{91.5}{\percent}$ of the data fall within the $\SI{95}{\percent}$ confidence interval, indicating strong consistency with the expected Gaussian model.

In more realistic conditions, where the system is subject to practical disturbances such as laser instability and environmental airflow fluctuations, the advantages of the EKF become more pronounced. As shown in Fig.~\ref{fig:nonsteady_noise}(a), poor laser stability introduces significant fluctuations in the measured light intensity (blue curve) that are unrelated to temperature variations. Conventional intensity-based inversion methods cannot distinguish whether intensity changes originate from the target sensing parameter or from laser instability. In contrast, the EKF accurately tracks the underlying signal, as indicated by the pink dashed curve.

We quantitatively evaluate the disturbance mitigation capability in temperature tracking by comparing three methods: a thermistor-based monitoring system, a conventional intensity-based inversion method using MZ interferometry, and the proposed EKF-based approach. To assess the temperature resolution of the three methods, the thermistor data were fitted using Eq.~\eqref{heating}, and the fitted curve was used as the reference for the mean temperature evolution $\overline{T}$. The residual for each method was then calculated as $T-\overline{T}$.  As shown in Fig. 5(b), the gray, blue, and pink curves correspond to the residuals of the thermistor-based method, the conventional intensity-based inversion method, and the EKF-based method, respectively. 

To further quantify the temperature resolution of each approach, the standard deviation (STD) of the residuals is calculated. The analysis indicates that the thermistor-based method exhibits the lowest resolution, with an STD of approximately $4\times10^{-3} \mathrm{K}$. The conventional intensity-based inversion method improves the resolution by about one order of magnitude to $2.28\times10^{-4} \mathrm{K}$. In comparison, the EKF achieves the highest resolution of $8.34\times10^{-5} \mathrm{K}$, representing an additional threefold improvement. These results validate that the proposed method possesses significant disturbance mitigation capabilities and high-resolution temperature tracking advantages. Although the true temperature is not directly accessible, these relative metrics clearly demonstrate that the EKF significantly enhances real-time temperature tracking and exhibits strong robustness against disturbances.

\section{Conclusion}
	
In this work, we develop an all-optical temperature sensing system based on a fiber interferometer. To enhance robustness against external disturbances, the EKF method is applied, in which the steady-state temperature is treated as a parameter estimation problem and the real-time temperature evolution is formulated as a state estimation problem. Simulation results show that this approach can accurately track temperature dynamics with the estimation error reduced to $2.21\times10^{-5} \text{K}$. In the experimental results, the EKF-based method outperforms both conventional intensity-based inversion methods and thermistor-based measurements, achieving the highest temperature resolution. Even under significant noise disturbances and nonstationary conditions, the system maintains a high resolution of $8.34\times10^{-5} \text{K}$, approximately three times better than that obtained from conventional intensity-based inversion methods.

These results indicate that the proposed system combines high resolution with strong robustness, demonstrating significant potential for high-precision dynamic temperature sensing applications. Its inherent advantages, including immunity to electromagnetic interference, compact structure, and low thermal capacity, further support its applicability across diverse measurement scenarios. In addition, under steady-state operation within the linear region of the response curve, the estimation framework can, in principle, be simplified to a standard KF, offering the potential for further performance improvement under optimal estimation conditions.

\begin{acknowledgments}
We thank supports from the Quantum Science and Technology-National Science and Technology Major Project (Grant No. 2024ZD0302200), and National Natural Science Foundation of China (NSFC) (Grants No.62522504, No.12374463). This work is also supported by the key R$\&$D Program of Zhejiang (2026C01004), Zhejiang Provincial Natural Science Foundation of China (ZCLQN25A0404) and Fundamental Research Funds for the Provincial Universities of Zhejiang (No. GK249909299001-002).
\end{acknowledgments}

	\bibliographystyle{unsrt}
	\bibliography{yang}
\end{document}